\documentclass[reprint, numerical, nofootinbib]{revtex4-1}
\pdfoutput=1
\usepackage{graphicx}   
\usepackage[skip=1pt,font={sf,footnotesize}]{caption}
\usepackage{color}      
\usepackage{subcaption}

\usepackage{hyperref}
\usepackage{amsmath}
\numberwithin{equation}{section}
\usepackage{amssymb}
\usepackage{verbatim}
\usepackage{float}
\usepackage{natbib}

\usepackage{chngcntr}
\counterwithout{equation}{section}

\usepackage{array} 
\usepackage{tabularx}
\usepackage{booktabs}
\usepackage[utf8]{inputenc}
\usepackage{bm}

\begin{document}
\title{Low-Dimensional Dynamics for Higher-Order Harmonic Globally Coupled Phase Oscillator Ensemble}
\author{Chen Chris Gong}
\email[]{cgong@uni-potsdam.de}

\affiliation{Institute of Physics and Astronomy, University of Potsdam, Karl-Liebknecht-Stra\ss e 32, 14476 Potsdam, Germany}

\author{Arkady Pikovsky}
\email[]{pikovsky@uni-potsdam.de}
\affiliation{Institute of Physics and Astronomy, University of Potsdam, Karl-Liebknecht-Stra\ss e 32, 14476 Potsdam, Germany}
\affiliation{Department of Control Theory, Nizhny Novgorod State University,
Gagarin Avenue 23, 606950 Nizhny Novgorod, Russia}

\date{\today}

\begin{abstract}
The Kuramoto model, despite its popularity as a mean-field theory for many synchronization phenomenon
of oscillatory systems, is limited to a first-order harmonic coupling of phases. For higher-order coupling, there only exists 
a low-dimensional theory in the thermodynamic limit. In this paper, we extend the formulation used by
Watanabe and Strogatz to obtain a low-dimensional description of a system of arbitrary size of identical oscillators coupled 
all-to-all via their higher-order modes. To demonstrate an application of the formulation, we use a 
second harmonic globally coupled model, with a mean-field equal to the square of the Kuramoto 
mean-field. This model is known to exhibit asymmetrical clustering in previous numerical 
studies. We try to explain the phenomenon of asymmetrical clustering using the analytical 
theory developed here, as well as discuss certain phenomena not observed at the level of first-order harmonic coupling.
\end{abstract}
\maketitle

\begin{quotation}
\end{quotation}

\section{Introduction} \label{intro}
Since its conception in 1975 by Yoshiki Kuramoto, the Kuramoto model of globally
coupled oscillators has been a standard tool used by diverse scientific 
communities, particularly within the fields of nonlinear dynamics, computational neuroscience and 
network science, to describe synchronization transition in ensembles of interacting oscillatory systems. 
It can be directly applied after justifiable phase reduction of the original system, 
and despite its
mathematical simplicity, captures the essential 
characteristics of synchronization phenomenon.%

The Kuramoto model is a model of all-to-all coupled ensemble of 
phase oscillators, with each oscillator represented by a scalar variable --- its phase. 
Inspired by the Ising model, Kuramoto's original
intention was to devise a similar model but for which there is
an analytically solvable transition
to synchronization, at least in the infinite system size limit 
(the thermodynamic limit) \cite{Kuramoto75, kura86}.
Kuramoto accomplished this by choosing the particular coupling function 
of two interacting oscillators to be proportional to the first harmonic 
function (i.e., sine or cosine) of the difference of two phases. %

Limiting the description of potentially complex periodic dynamics to a scalar phase for each interacting
sub-unit may appear to be highly restrictive at a first glance. However,
it was shown that a phase oscillator model such as Kuramoto model
approximates the long-term behavior of any ensemble of 
interacting oscillatory systems, so long as the coupling 
is weak and the sub-units are nearly identical \cite{Kura84}. The
oscillators are said to be weakly coupled if their mutual perturbations 
via their interactions are small (1) when compared to the characteristic strong stability of the
oscillators amplitudes, and (2) when compared to their 
intrinsic natural frequencies (the changes in the period are small compared to the periods). 
There are many examples of reduction of a realistic oscillatory system to the Kuramoto 
phase oscillator model, such as for Josephson Junctions \cite{JJ}, 
atomic recoil lasers \cite{Politi08}, functional connectivity of the human brain \cite{Cabral11, Petkoski18}
and in \emph{Caenorhabditis elegans} \cite{Moreira}, neuronal oscillations \cite{Daffertshofer10, Politi15, Pazo18}, power networks
and smart grid \cite{Bullo12, Filatrella2008}. %

Despite the canonical status of the Kuramoto model, many oscillators interact with each other 
beyond the simple picture of first harmonic coupling. 
Recently there has been an increasing interest in second harmonic coupling functions
and other forms of coupling via higher-order modes ---
such models of globally coupled phase oscillators are often called Kuramoto-Daido
models~\cite{Daido92, Skardal, Komarov13, higherorder_chimera, Zhang16, Yang17}. 
There are indeed many experimental situations where the second harmonic coupling is large and even dominates over
the first harmonic \cite{Czolczynski13, Goldobin11, Goldobin13, Kiss05, Kiss06}. Second harmonic
coupling can imply non-pairwise connection, which have been shown to exhibit multistability 
and chaos \cite{Tanaka11, maxim, Bick16, Skardal19}.
Higher order mode coupling usually means that a coupling function $\Gamma(\varphi_k-\varphi_j)$
between each pair of oscillators is a generic $2\pi$-periodic
function of the phase difference $\varphi_k-\varphi_j$, containing a few or many harmonics. 
Phenomenologically, when higher harmonics are dominant in an interaction, 
the synchronous state of the system is characterized by the formation of 
multiple synchronized groups (or ``clusters'') of oscillators, 
each with a common phase \cite{Hansel93}. This differs from the cases where only the first
harmonic exists, which can result in at most one cluster. %

A remarkable feature of the first-order harmonic global interaction, 
is that it allows for a low-dimensional reduction \cite{Ott08, ws94},
i.e., a two- or three-dimensional dynamics suffices at describing an N-body interaction.
Similarly, there is also a hidden low dimensional dynamics for a
pure higher-order coupling in the thermodynamic limit \cite{Skardal}, which was
shown using a similar method as the one employed by Ott and Antonsen \cite{Ott08}
for the first harmonic coupling.
In this study, we concern ourselves with another dimension-reducing technique 
that was developed earlier
than the Ott-Antonsen (OA) theory, namely the Watanabe-Strogatz (WS) 
theory \cite{ws94, PR08, PR15, Chen_2017}.
Unlike the OA theory, the WS theory does not need a special ansatz 
and can also be applied to a finite-sized ensemble, however, it is 
restricted to oscillators with identical natural frequencies that are
identically driven. 

In the following sections, we show that
the WS theory indeed can be extended to pure higher-order models. In Section \ref{sec:Prob}, 
we first introduce the general model of pure higher-order harmonic coupling.
In Section \ref{sec:theory}, we review the WS theory for a general first-order harmonic coupling 
of the Kuramoto-Sakaguchi kind, then extend it to pure higher-orders. Last, in Section \ref{sec:num},
we apply the extended WS theory to a non-trivial second harmonic model exhibiting asymmetrical clustering, 
and conduct numerical simulation of its low-dimensional 
WS equations. 
We also find that under certain special initial conditions, such a second harmonic model 
could exhibit de-coherence under attractive coupling, which is not found in first-harmonic models.
\section{Formulation of the Model} \label{sec:Prob}
We study a population of $N$ identical phase oscillators with phases $\{\varphi_j\}$, $j = 1, 2, \ldots, N$, subject
to a global coupling.
Here, unlike in the standard Kuramoto-Sakaguchi model \cite{kura86}, the coupling term is purely 
of an arbitrary higher-order $l$ ($l \geq 2$),
\begin{equation} \label{eq:gencoup} 
\dot \varphi_{j} = \omega(t) + \operatorname{Im}[ H(t) e^{-il\varphi_{j}}] ~, 
\end{equation}
where $\omega(t)$ and $H(t)$ are arbitrary scalar 
and complex functions, respectively. 
When $\omega$ is constant, it represents the identical natural 
frequency of the oscillators. While in the problem formulation and theory derivation we write generically $\omega(t)$,
 in the numerical part we use a constant $\omega$,
to be comparable to previous numerical studies in the literature.
$H(t)$ represents an arbitrary complex forcing term,
which can be dependent or independent of the phases $\{\varphi_j\}$, deterministic or stochastic, and also 
can be external time-dependent forces. The latter case is not considered in this paper; see 
Ref.~\cite{Multicluster} for exploration of external driving within the scope of the reduced WS theory
for the first-order coupling.

Global coupling (a.k.a. ``all-to-all'' coupling) of the oscillators corresponds to the case where $H(t)$
depends on the Kuramoto-Daido order parameters (mean-fields of the higher harmonics of phases)
\[
Z_m=\frac{1}{N}\sum_j e^{im\varphi_j}\;.
\]
For simplicity, in the rest of the paper we use $Z_1$ and $Z$ interchangeably to denote the Kuramoto order parameter,
which is also the first Kuramoto-Daido order parameter. 

The simplest example of high-order coupling of type Eq. \eqref{eq:gencoup} is a model of identical phase oscillators 
globally coupled via the second harmonic coupling function 
of their phase differences:
\begin{equation} \label{eq:Daido}
\dot \varphi_{j} = \omega + \frac{1}{N}\sum\limits_{k = 1}^{N}\sin(2\varphi_{k} - 2\varphi_{j} + \gamma)~=\omega+\text{Im}(Z_2e^{i\gamma}e^{-2i\varphi_j})~, 
\end{equation}
where $\gamma$ is the phase shift parameter, tuning the nature of the coupling between various degrees of 
attractiveness or repulsiveness. Here the global forcing term $H(t)$ 
is just the second-order Kuramoto-Daido mean-field $Z_2$ rotated by the phase shift $\gamma$.

This system is trivial to solve due to its similarity with the Kuramoto model, with
phases $\varphi$ now replaced by $2\varphi$ and everything else stays the same (Ref. \cite{Delabays19} has shown
that they are fully equivalent). 
Below we focus on  
more complex models, where $H(t)$ is a generic function of order parameters, which satisfies the phase shift invariance 
property (i.e., under $\varphi\to\varphi+\text{const.}$ the dynamics is the same).  
In particular, the complex forcing can take any form such as $(Z_{q})^m (Z_p^*)^n$, with $mq-pn=l$,
or a combination of these terms. So for example, for $l=2$ one can have
$H(t) \sim Z_2$ like in Eq. \eqref{eq:Daido}, but also $H(t)\sim Z^2$ like in Ref.~\cite{maxim}, or, e.g.,
$H(t) \sim Z_4 Z_2^{*}$.

\section{Theory} \label{sec:theory}

\subsection{Watanabe-Strogatz theory for Kuramoto-Sakaguchi model} \label{sec:wst_1}
Before delving into the treatment of higher-order harmonic coupling using WS theory, 
we review first the original formulation which deals with the first-order harmonic coupling, 
i.e., the Kuramoto-Sakaguchi model. In 1994, in modelling arrays of $N$ identical 
overdamped Josephson junctions, Watanabe and Strogatz~\cite{ws94} showed that such a system
 has hidden low-dimensional dynamics, for which $N-3$ constants of motion 
exists. This theory, which we shall call the WS theory, is applicable to $N$-dimensional 
dynamics of a system of identically driven identical phase oscillators described by
\begin{equation}
\dot\varphi_j =\omega(t)+\textrm{Im}[H(t) e^{-{i} \varphi_j}],\quad j=1,\ldots,N~,
\label{eq:gws}
\end{equation}
where $\omega(t)$ and $H(t)$ are arbitrary real and complex-valued functions of time, respectively. 
When $\omega$ is a constant, it represents the common natural frequencies of the oscillators. 
When $H(t) \sim Z$,
this system corresponds to the Kuramoto-Sakaguchi model of globally coupled identical oscillators. 

A coordinate transformation $\mathcal{M}_1$ which is called the M{\"o}bius transformation is central to
the WS theory (see Refs.~\cite{ws94,Marvel-Mirollo-Strogatz-09} for a detailed presentation). 
$\mathcal{M}_1$ formally belongs to the class of M{\"o}bius maps (or M{\"o}bius group action), 
which is a type of fractional linear transformation, mapping the unit circle 
in the complex plane to itself in a one-to-one way. Explicitly, the time-dependent M{\"o}bius 
transformation and its inverse\footnote{Distinguishing forward and inverse transformations is rather arbitrary; here we just
use one possible formulation.} can be written as
\begin{align} \label{eq:moebtrafo}
\mathcal{M}_1&: \psi_{j} \rightarrow \varphi_{j}(t), 
\hspace{3mm} e^{{i} \varphi_{j}(t)} = \frac{ z(t) + e^{{i}(\psi_{j}+\alpha(t))} }{ 1+z^*(t) e^{{i}(\psi_{j}+\alpha(t))} } ~, \\
\mathcal{M}_1^{-1}&: \varphi_{j}(t) \rightarrow \psi_{j}, \hspace{3mm} e^{i \psi_{j}} = e^{-i\alpha}
\frac{ z(t) - e^{{i}\varphi_{j}(t)} }{ z^*(t) e^{{i}\varphi_{j}(t)} - 1} ~.
\label{eq:invmt}
\end{align}

Here $\{\varphi_{j}\}$ are the phases of the oscillators, complex parameter $z(t)$ satisfies 
$|z(t)| \leq 1$, and the parameter $\alpha(t)$ is a rotation angle.
If the phases evolve according to \eqref{eq:gws} and the WS parameters
$z$ and $\alpha$ evolve according to
\begin{align} \label{eq:WS}
\dot z &= {i}\omega(t) z + \frac{1}{2} H(t) - \frac{1}{2} H^{*}(t) z^{2},  \\
\dot \alpha &= \omega(t) + \textrm{Im}[z^{*} H(t)] ~,\nonumber
\end{align}
then the transformed phases $\psi_j=\mathcal{M}_1^{-1}(\varphi_j)$ are conserved quantities 
(``constants of motion''). Thus, WS theory implies partial integrability
of the system of identical oscillators.
Equation~\eqref{eq:WS} can be shown to be
a Riccati equation, and its integrability follows from the transformation of the
Riccati equation to a linear form \cite{Goebel95, Chen_Thesis}. 

\begin{figure}[t!]
\includegraphics[width=0.85\columnwidth]{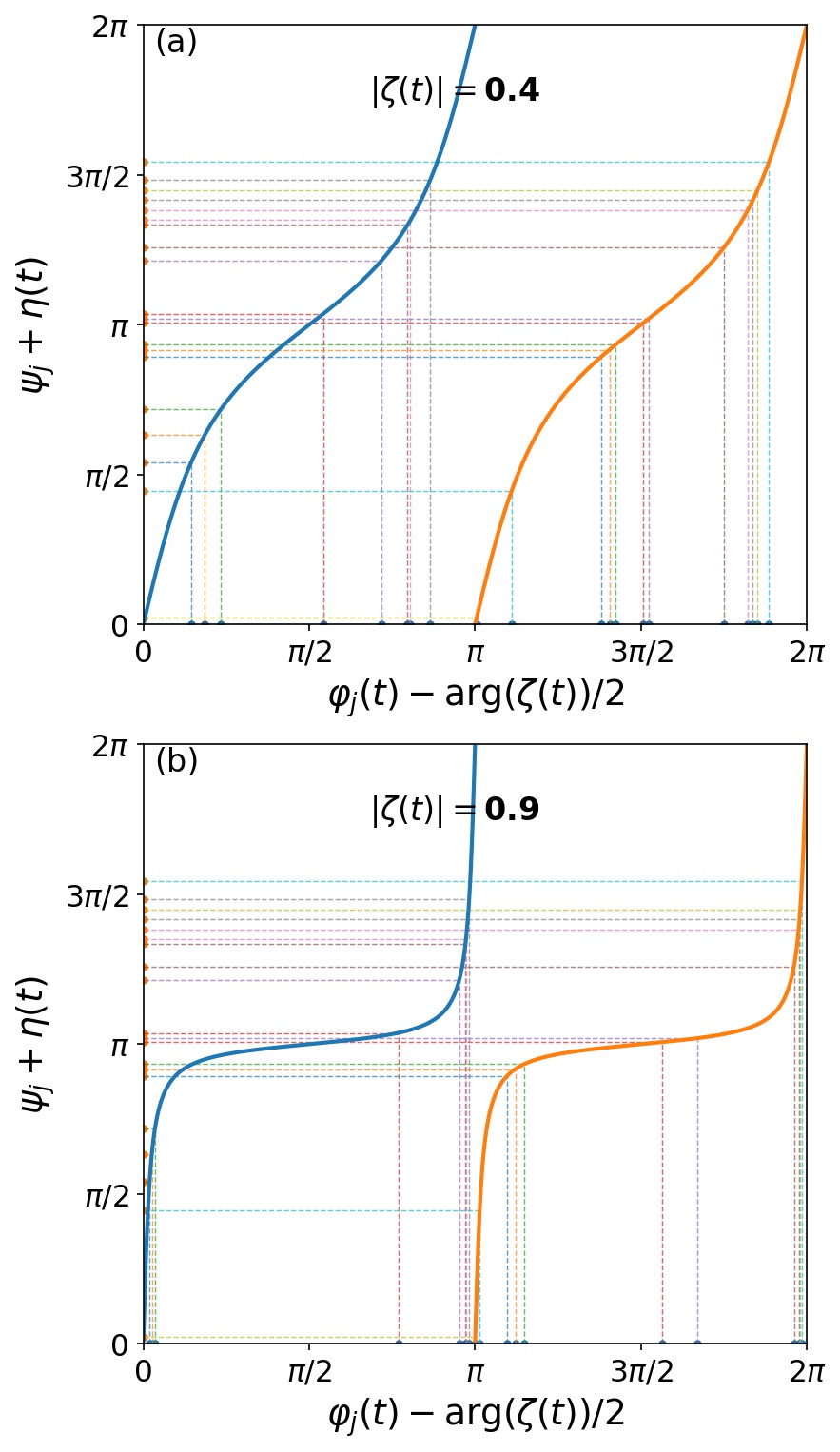}
	\caption{The M\"obius transformation $\varphi_{j}(t) \leftrightarrow \psi_{j}$ for the second harmonic coupling
	is visualised for two parameter values of $|\zeta(t)|$. Horizontal axis: phases $\varphi_j$ 
	are shifted by the parameter $\arg(\zeta(t))/2$; Vertical axis:
	constants $\psi_j$ are shifted by the parameter $\eta(t)$.  Transforming the same set of constants 
	to the phases
	results in a more spread out set of phases (panel (a))
	for small values of $|\zeta|$, 	and in a more clustered state of phases for $|\zeta|$ close to one (panel (b)).
	The two branches of the mapping illustrate the non-uniqueness of $\mathcal{M}_2(t)$.}
	\label{fig:Mtrafo}
\end{figure}

Under the M{\"o}bius transform Eq. \eqref{eq:moebtrafo}, constants $\psi_j$ are rotated by 
the angle $\alpha$ and then contracted along the circle into the direction of $\arg[z(t)]$,
the degree of contraction controlled by $|z(t)|$ (see also
a visualization of second harmonic example in Fig. \ref{fig:Mtrafo}). In fact, akin to Kuramoto 
order parameter $|Z|$, $|z|$ can typically be used 
as a measure of synchronization, since both parameters become equal to unity at full synchrony.
 
Because we have introduced three extra parameters via the M\"obius transform, 
to make the M\"obius transform unique, we must impose the same number of conditions 
on the new system Eq. \eqref{eq:WS}. We have the choice of either imposing three conditions 
on the constants of motion, or, we can impose conditions
on the initial values of the parameters themselves. 
The conditions themselves are rather arbitrary. In practice however, there
are a number of ways of choosing conditions such that the system evolve 
more ``naturally''. For the WS reformulation of higher-order coupled system 
(see Sec. \ref{sec:num}), we focus on the latter option, namely, imposing conditions on the parameters'
initial values. 

\subsection{Generalization of the WS theory to a coupling via higher harmonics} \label{sec:wst_l}
Here we generalize the WS approach outlined above to coupling via higher harmonics, using 
derivation extremely similar to those outlined in Ref.~\cite{PR15}. Due to the algebraic similarity,
we only sketch out a general idea, and leave the details to be inferred from Ref.~\cite{PR15}.

$N$ phase oscillators coupled via higher-order harmonics obey the general equations of 
motion Eq. \eqref{eq:gencoup}. It can be rewritten as
\begin{equation} \begin{aligned} \label{eq:gencoup2}
\frac{d }{d t} (e^{i l \varphi_{j}}) = i l  e^{i l \varphi_{j}} \omega(t) + \frac{l}{2} \left(H(t) - H^{*}(t) e^{ 2 i l \varphi_{j} } \right)~.
\end{aligned} \end{equation}
We transform the phases $\varphi_j$ into phases $\vartheta_j$
via
\begin{equation}
e^{ i l\varphi_{j}} = \frac{ \zeta + e^{i\vartheta_{j}} }{ 1 + \zeta^{*} e^{i\vartheta_{j}} } ~,
\end{equation}
with an additional complex parameter $\zeta$. 
Equations \eqref{eq:gencoup2} can be transformed in terms of $\{\vartheta_{j}\}$, $\zeta$ and 
their time derivatives $\{\dot \vartheta_{j}\}$ and $\dot{\zeta}$. Going through a similar
procedure of picking out terms in the orders of 
$e^{i \vartheta_{j}}$ as done in Ref.~\cite{PR15}, we obtain
\begin{equation} \label{eq:ws-l}
\begin{cases} 
\dot{\zeta} = l\left[\imath\omega(t) \zeta + \frac{1}{2}H(t) - \frac{1}{2}H^*(t) \zeta^2\right] \;,\\
\dot {\vartheta}_j = l\left\{\omega(t) + \operatorname{Im} [H(t) \zeta^{*}] \right\} ~,
\end{cases}
\end{equation}
which satisfy all $N$ transformed equations, and hence also the $N$ 
original equations \eqref{eq:gencoup2}. We notice that the
right-hand side of the second equation of Eqs. \eqref{eq:ws-l} is independent of $j$, indicating that 
 all the angles $\{\vartheta_j\}$ rotate at the same speed. 
Therefore, we can create a new time-dependent parameter $\eta$ which has the same rotational speed
as $\{\vartheta_j\}$, $\dot {\eta} = \dot {\vartheta}_j $, and define 
\begin{equation}
\eta(t):= \vartheta_{j}(t) - \psi_{j}\;,
\end{equation}
where $\{\psi_{j}\}$  are the constants of motion. 

We thus come to the M\"{o}bius transformation from constants $\psi_j$ to phases $\varphi_j$
\begin{align}\label{eq:mobiusPR_higher}
\mathcal{M}_l : \psi_{j} \rightarrow \varphi_{j}(t), \hspace{3mm} e^{ i l\varphi_{j}(t)}& = 
\frac{ \zeta(t) + e^{i\vartheta_{j}(t)} }{ 1 + \zeta^{*}(t) e^{i\vartheta_{j}(t)} } \\
&= 
\frac{ \zeta(t) + e^{i[\psi_{j} + \eta(t)]} }{ 1 + \zeta^{*}(t) e^{i[\psi_{j} + \eta(t)]} }~, 
\end{align}
depending on the time-dependent WS variables $\zeta(t),\eta(t)$.
 The inverse M\"{o}bius transformation for higher-order coupling is
\begin{equation}\label{eq:mobiusPR_higher_inverse_psi}
\mathcal{M}_l^{-1} : \varphi_{j}(t) \rightarrow \psi_{j}, \hspace{3mm} e^{i \psi_{j}} = e^{-i\eta(t)}
\frac{ \zeta(t) - e^{{i}l\varphi_{j}(t)} }{ \zeta^*(t) e^{{i}l\varphi_{j}(t)} - 1} ~.
\end{equation}

Compare Eqs. \eqref{eq:mobiusPR_higher} and \eqref{eq:mobiusPR_higher_inverse_psi}
to the transform for a first-order coupling Eqs.\eqref{eq:moebtrafo} and \eqref{eq:invmt},
only the original phases are multiplied with the order of coupling $l$, otherwise the form
of the transform stays the same. 
Comparing the WS equations for the first-order coupling Eq.~\eqref{eq:WS} with Eq.~\eqref{eq:ws-l}, 
we find that the equations for pure higher harmonic (or ``$l$-harmonic'') coupling are merely 
multiplied by the factor $l$ on the right-hand side.

We can write the equations for the three WS parameters
Eq.\eqref{eq:ws-l} in terms of dot and cross 
products of $H(t)$ and $\zeta$ in the complex plane ($\zeta = \rho \exp(i \Phi)$, $\rho \ne 0$):
\begin{equation} \label{eq:WS1_rho} \begin{aligned}
\dot{\rho} & = l \frac{1- \rho^2}{2\rho} H(t) \cdot \zeta, \\
\dot{\Phi} & = l [\omega(t) + \frac{1+ \rho^2}{2\rho^2} H(t) \times \zeta],\\
\dot {\eta} & = l [\omega(t) + H(t) \times \zeta],
\end{aligned} \end{equation}
where parameter $\Phi$ evolves according to $H(t) \times \zeta$, similar to
a torque experienced by an object with a magnetic moment under a magnetic field.
For different $H(t)$ it is as if the same magnetic moment, denoted by the higher-order
WS parameter $\zeta$, moves under a different magnetic field. 

\subsection{Numerical simulation of the dynamics in the WS variables}
At first glance, Eqs.~\eqref{eq:WS1_rho} present an enormous simplification
compared to the original model~\eqref{eq:gws}, as the number of equations is reduced from $N$ to $3$.
However, the difficulty in numerical simulation
of the WS equations is that the coupling term $H(t)$ is typically expressed in terms of the original phases
via the Kuramoto-Daido order parameters, and not in terms of the WS variables and the
constants of motion. Therefore, for each calculation of right-hand side in Eq.~\eqref{eq:WS1_rho} one has
to perform transformation Eq.~\eqref{eq:mobiusPR_higher}. If the coupling contains
only order parameters $Z_{m*l}$ with integer $m$, then only quantities $e^{il\varphi_j}$
are needed to compute the coupling term and no transformation step is needed. 
However, if other order parameters have to be calculated, then
one needs to know phases $\varphi_j$, and they are not uniquely defined through quantities $e^{il\varphi_j}$.
Indeed, one value of a constant $\psi$ maps to $l$ values of the phase variable:
$\varphi/l + 2 n \pi/l$, where $n = 0, 1, \ldots, l-1$. To choose a proper value, 
one can use continuity of the
dynamics of the phases $\varphi$ in time. 
This means, the proper value of the phase
at time instant $t+\Delta t$ is the value closest to that at the previous step $\varphi(t)$, for small $\Delta t$.
In numerical implementations without intermediate steps, like Euler or Adams-Bashforth
schemes for solving ordinary differential equations, this check is simple. In Runge-Kutta-type schemes,
one should take care that also at intermediate calculations of the right-hand side of equations
inside a Runge-Kutta step, the proper phase is extracted from the 
transformation Eq.~\eqref{eq:mobiusPR_higher}.

\subsection{Basins of attraction for clusters} \label{sec:wst}

The WS theory implies that a system of globally coupled identical 
oscillators with an $l$-harmonic coupling 
can evolve to at most $l$ clusters at any point in time. Indeed, if the initial phase distribution 
has no clusters, then all the constants of motion
$\psi$ are different. Then, for any $|\zeta|<1$, all the phases are different as well. 
The only way for clusters to form is $|\zeta|\to 1$ under attractive coupling.
 
For attractive $l$-harmonic coupling, in general it is expected that eventually
the phases form $l$ clusters, i.e.,,  $l$ distinct attractive subgroups of oscillators (there are special
initial states for which this is not true, see discussion in Sec.~\ref{sec:dmf} below
about the solutions in which $|\zeta|$ does not grow). 
Thus, the circle is divided in $l$ basins of attraction of these clusters. 
 The boundaries of these basins of attraction are hence special points 
of the collective motion, since they will not be synchronized to any final cluster, 
and can be described as ``unsychronizable'' (``solitary states'' in the terminology of Ref.~\cite{solitarystate}). 
Because basins evolve in time, the boundaries are unstable trajectories
of the dynamics on the unit circle. Below we relate
these boundaries to the mathematical singularity occurring in the WS 
formulation of the system, specifically, to the pole in the 
M{\"o}bius transformation Eq.~\eqref{eq:mobiusPR_higher}. 

Because basin boundaries are not trajectories of real oscillator phases,
we have to consider the transformation Eq.~\eqref{eq:mobiusPR_higher} for all possible values of $\psi$.
One can see that this transformation becomes singular at the limit $|\zeta|\to 1$. For $|\zeta|=1$,
all values of $\psi$ are mapped to the cluster states  $\varphi = \Phi/l + 2n\pi/l$, where $n = 0, \ldots, l-1$,
except for the singular value $\tilde\psi=\Phi+\pi-\eta$, where $\Phi=\arg(\zeta)$ as defined in Section \ref{sec:wst_l}~. 
This singular constant is
mapped via Eq.~\eqref{eq:mobiusPR_higher} (at $|\zeta| \rightarrow 1^-$ when the map is not singular) 
to the basin boundaries at the end of the evolution $t\to\infty$: 
$\tilde\varphi = \Phi/l + (2n+1) \pi/l$, where $n = 0, \ldots, l-1$.

In a particular case under $l=2$ to be explored numerically below, we have two such basin boundary trajectories.
At the end of the evolution, at $t\to\infty$, where clusters are formed 
and $|\zeta(\infty)| = 1$, these are points $\tilde\varphi_1(\infty) = \Phi(\infty)/2 + \pi/2$ and 
$\tilde\varphi_2(\infty) = \Phi(\infty)/2 + 3\pi/2$. To find these boundaries at all times,
and in particular at the initial moment in time, one can trace these states back in time, but even that is not necessary. 
In fact, to find $\tilde\varphi_{1,2}(t)$, it is sufficient to know the singular value of the constant $\tilde\psi$
at the final stage of the evolution: $\tilde\psi=\Phi(\infty)-\eta(\infty)+\pi$. Then, for each $0\leq t<\infty$, the basins can 
be calculated according to the transformation Eq.~\eqref{eq:mobiusPR_higher}:
\begin{equation}
e^{i2\tilde\varphi(t)}=\frac{\zeta(t)+\exp[i(\tilde\psi+\eta(t))]}{1+\zeta^*(t)\exp[i(\tilde\psi+\eta(t))]}\;.
\label{eq:bb}
\end{equation}
This expression shows, that at each moment of time the basin boundaries can be obtained via $\tilde\psi$ transformed
by parameters  $\zeta(t)$ and $\eta(t)$. The expression also tells us that the sizes of the basins are equal (in the case of $l=2$,
the sizes are $\pi$). However, the positions of the basins depend on the final point of integration of both WS
variables $\zeta(t=\infty)$ and $\eta(t=\infty)$: thus to find them one first has to perform integration up to large 
enough time, and only after that formula Eq.~\eqref{eq:bb} is applicable.  Below we will also discuss an approximate 
way to define these boundaries solely from the initial state, and will see that 
it does not provide an exact prediction of the clustering.

\section{Numerics} \label{sec:num}
In this section we numerically simulate a particular model coupled via
second-order harmonic which has been studied in previous literature. 
The approach can be generalized to arbitrary higher-order of phase coupling, 
and we offer one example with the fifth order
coupling in the \hyperref[Appendix]{Appendix} for completeness. 

\subsection{Higher order harmonic coupling example: $Z^2$ mean-field} \label{sec:Zsqr}
As discussed above, for $l=2$ a coupling scheme via the second-order Kuramoto-Daido order
parameter $Z_2$ is trivial, because it can be reduced to the standard Kuramoto model.
It appears that a simple nontrivial example is a coupling via the square of the first-order mean-field, i.e., 
$H(t) = Z^2$. This model has appeared in previous literature \cite{maxim}, where
an ensemble of identical phases at steady state is always found to exhibit a curious strictly non-symmetric 
two-cluster distribution (or ``asymmetrical clustering'' in literature), starting from phases
drawn randomly from a uniform distribution on the circle. It is ``strictly'' asymmetric because one 
cluster always contains more oscillators than the other in the final state. It is curious because
the apparent asymmetry in the final distribution arise deterministically from 
identical oscillators identically driven, with uniform random initial conditions. To  
further study this distribution, we use the extended WS formulation above and its prediction of the boundaries of
the two basins of attractions to partially explain the source of this apparent symmetry breaking. 

The equations for $Z^2$-mean-field model of identical oscillators can be written as the following
\begin{equation}
\dot \varphi_{j} =|Z|^2\sin(2\arg(Z)-2\varphi_j),\quad Z=\frac{1}{N}\sum_j e^{{i} \varphi_j}\;,\label{eq:ZsqrODE0}
\end{equation}
or
\begin{equation}
\dot \varphi_{j} = \frac{1}{N^{2}} \sum\limits^N_{k=1} \sum\limits^N_{m=1} \sin(\varphi_{k} + \varphi_{m} - 2 \varphi_{j}) ~,
\label{eq:ZsqrODE}
\end{equation}
which corresponds to Eq. \eqref{eq:gencoup} with $l = 2$ and $H(t) = Z^2$. 
Moreover, we assume the natural frequency $\omega$ 
to be a constant and fix its value to zero (one
can accomplish this by choosing a rotating reference frame).

Since we can rescale time, we have set the coupling strength to 1 without loss of generality. 
The elementary coupling between phases Eq.~\eqref{eq:ZsqrODE}, unlike the $Z_2$-mean-field model Eq.~\eqref{eq:Daido},
involves now a triplet of oscillators indexed by $m, k$, and $j$. This corresponds to a hypernetwork topological 
connection between the oscillators, where three nodes jointly form a coupling connection, as opposed to a 
normal network where only two nodes are needed for a coupling connection.
This hypernetwork model may play an important role in neuronal coupling \cite{Petri14, Giusti16, Sizemore18}.

As discussed above, for $t\to\infty$,
 two clusters will form with some constant final value of $Z$, one with the
 phase of the mean-field $\arg(Z)$ and the other one shifted by $\pi$: 
 $\arg(Z) + \pi$, as can be easily found from Eq.~\eqref{eq:ZsqrODE0} by equating the right-hand side to zero.
A simple metric for describing the distribution of the phases among the clusters is $R := |Z|$, the Kuramoto order parameter amplitude. 
It relates to the population of one of the clusters by $R = |2 N_{1}/N - 1|$, where $N_1$ is the number of 
oscillators in one of the two clusters. When $R = 0$, the two clusters have equal size. When $R=1$,
all the oscillators are in one cluster. 


\subsection{Integration of the WS equations for the $Z^2$-mean-field model} \label{sec:WSZsqr}

Before we carry out numerical integration of WS equations, we introduce a method of visualizing 
the basins. As discussed above, one needs to follow the evolution not only for the set of coupled
oscillators, but for all possible values of phases that can be mapped to the space of the 
constants $\psi$. Equivalently, we can use 
Eq. \eqref{eq:ZsqrODE0}, and unidirectionally couple an arbitrary number of oscillators to the field.
These oscillators, which we denote $\theta$ as passive traces, are influenced by but
do not contribute to the global field which depends on the ``active'' phases $\varphi_j$ only
\begin{equation} \label{eq:ZsqrODE_passive}
\dot \theta = \operatorname{Im}[ Z^2 e^{-i2\theta}]  ~,
\end{equation}
where the mean-field is defined in Eq. \eqref{eq:ZsqrODE0}. 
Variable of a tracer $\theta$ is not indexed since we can use any number of them
and they take on any value between 0 and $2\pi$. 

Introducing passive oscillators gives us the advantage of visualizing the 
field on the entire circle, because we can place them anywhere on the circle to ``test'' the 
strength of the field, and not just at those places where the active oscillators
happen to be. In this sense they are analogous to the fluid tracers in hydrodynamical 
simulations or experiments. It will make the motion of the points on the circle under the field
obvious to the eye, especially those near the unstable points. 

For the $Z^2$ mean-field model Eq.~\eqref{eq:ZsqrODE0}, the WS parameters obey
\begin{equation} \label{eq:wsii_realimag}
\begin{cases} 
\dot{\zeta} =  Z^2(t) - [Z^*(t)]^2 \zeta^2 \\
\dot {\eta} = 2\operatorname{Im} [Z^2(t) \zeta^{*}] ~.
\end{cases}
\end{equation}

Initial values of the WS parameters in our numerical simulation
are chosen as $\zeta(0) = Z^2(0)$ and $\eta(0) = 0$. Under such an initial condition,
the second WS equation $\dot \eta = 0$ at $t= 0$, therefore,
it can be considered as a natural initial condition, 
although it is not the only reasonable one.
For instance, previous literature \cite{ws94} has given two initial conditions as options.
One is the ``identity conversion'', 
with the introduced WS parameters all set to 0: $|\zeta(0)| = 0$, $\arg[\zeta(0)] = 0$ and $\eta(0) = 0$, which corresponds to when $\mathcal{M}_1$
is just the identity operator at $t=0$. The other is the ``incoherent state'', 
which corresponds to when the constants of motion is maximally incoherently distributed,
i.e., choose $\zeta(0)$ and $\eta(0)$ such that $\langle \exp(i\psi_j) \rangle = 0$
(if no majority cluster exists). 
``Identity conversion'' was deemed unsuitable because even with different initial sets of phases,
the WS parameters start at the same point in the three-dimensional phase space. 
However, our chosen initial condition for the parameter,
$\zeta(0) = H(0)$, does depend on the initial phases. 
This initial condition is also more suited to the complex representation of the WS system, as opposed to 
the three real equations in Ref.~\cite{ws94} or like Eq. \eqref{eq:WS1_rho}, 
since $\rho(0):=|\zeta(0)| = 0$ is a singularity there, 
and $\arg[\zeta(0)]$ would be undefined.
For clarity, we define explicitly the argument of 
$\zeta$: $\Phi = \arg(\zeta)$ as before.
 
\begin{figure}
 \includegraphics[width=.45\textwidth]{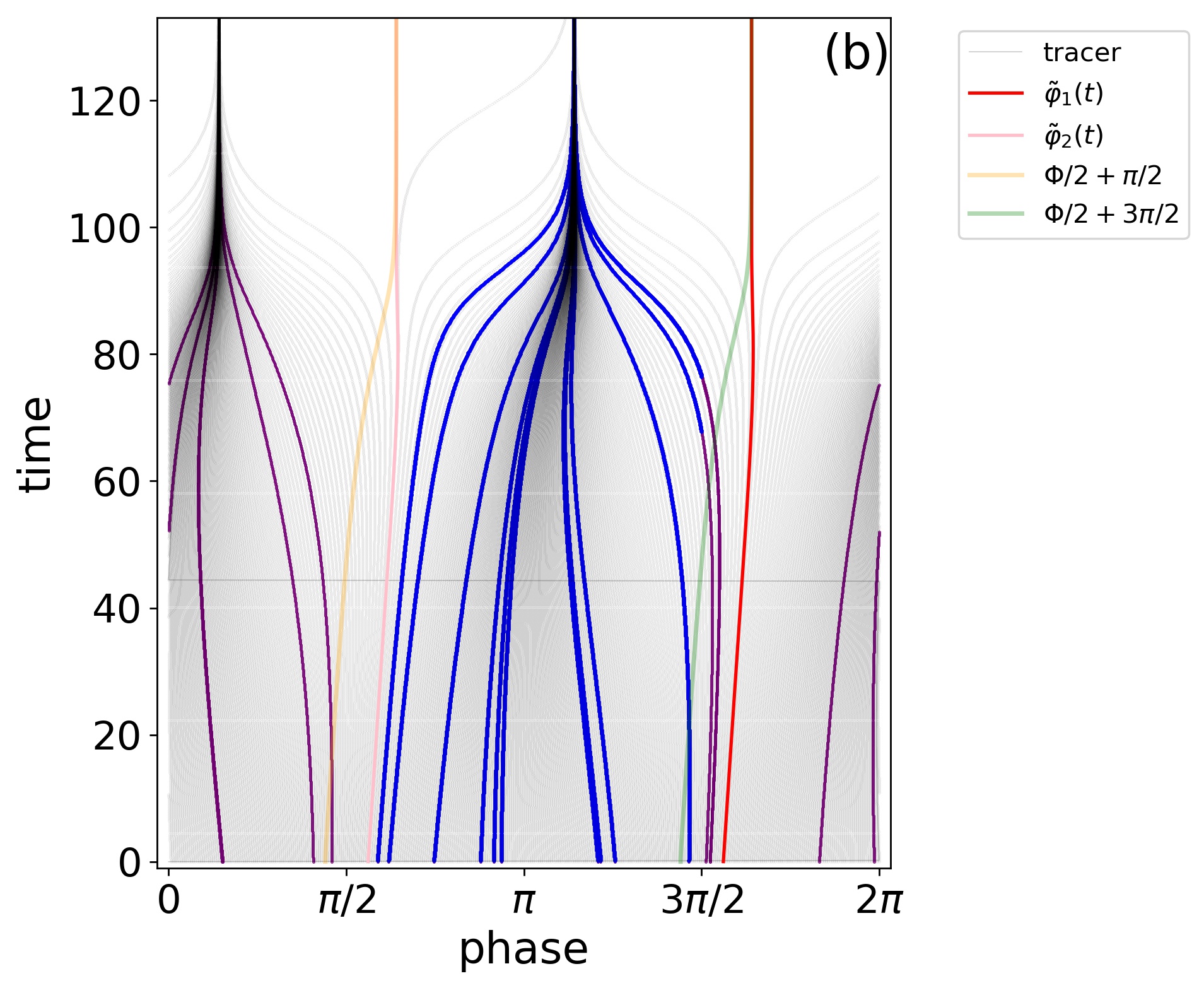}
 \includegraphics[width=.45\textwidth]{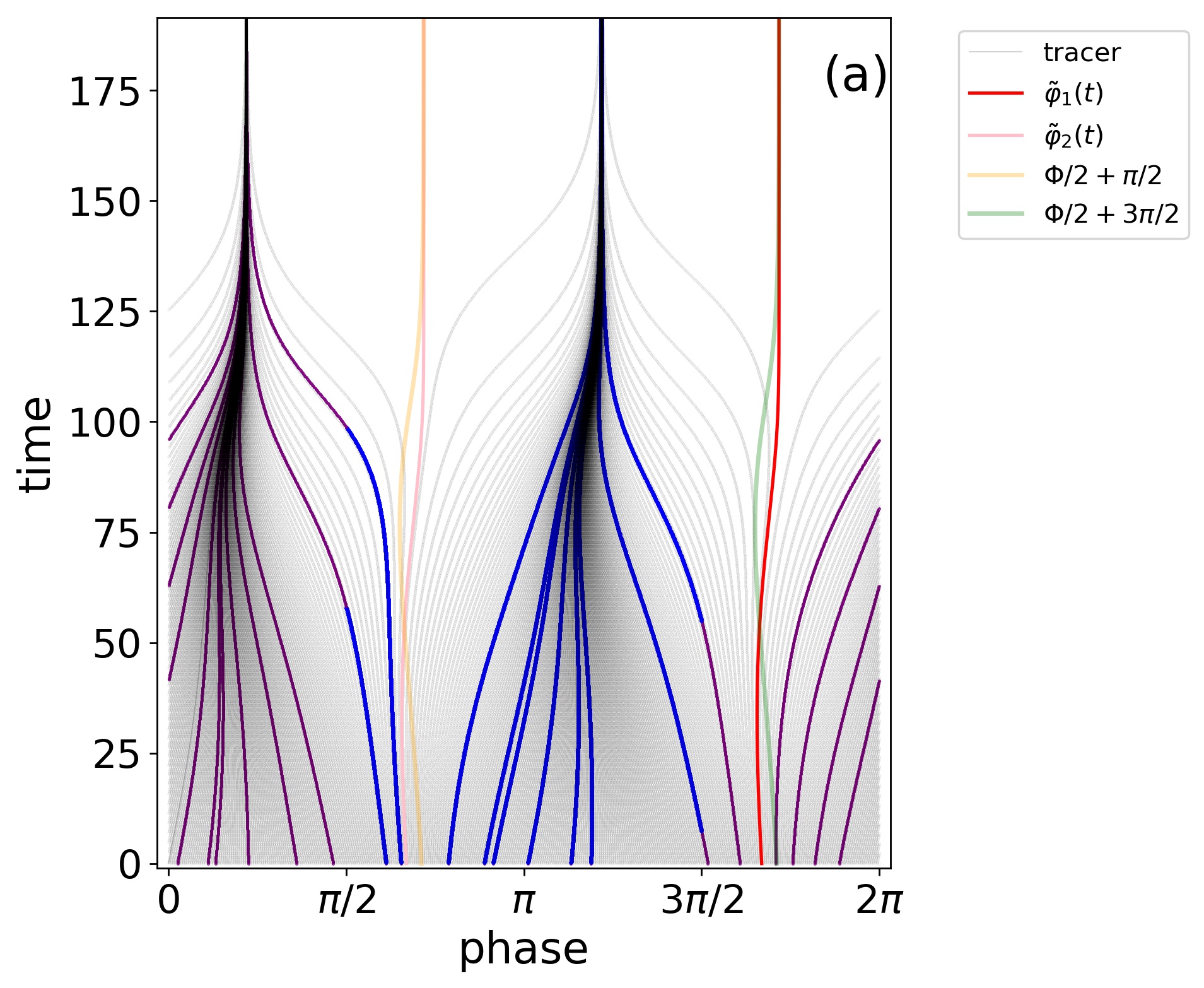}
 \caption{Euler integration with $h = 0.01$ of the WS equations
 \eqref{eq:wsii_realimag} for the $Z^2$-mean-field model. They are simulated for 
 two sets of random initial conditions (phases are randomly drawn from uniform distribution from
 0 to $2\pi$). Integration is carried out until two synchronized clusters are formed. 
 Gray lines are the tracers $\theta$ (Eq. \ref{eq:ZsqrODE_passive}), 
 which are uniformly spaced initially on a circle,
 and passively coupled to the global field of the active phases. 
 The flow of $20$ active phases $\varphi_j$ are marked by purple or blue. 
 Purple indicates if at time 
 $t$, the phase $\varphi_j$ transformed back from the constant $\psi_j$
 does not need to be added $\pi$, and blue indicates if it does, 
 to ensure continuity of the flow of the phases. 
 Trajectories of WS parameter $\Phi(t)/2 + \pi/2$ and $\Phi(t)/2 - \pi/2$
 are in orange and green.
 Pink and red lines are the trajectories of two 
 tracers which end up at singular points $\tilde\varphi_1(\infty)$ and 
 $\tilde\varphi_2(\infty)$ (as discussed in Sec. \ref{sec:wst}). 
 These unstable trajectories are computed via Eq.~\eqref{eq:bb} from 
 the singular constant $\tilde\psi$ and the saved values of $\zeta(t)$ and $\eta(t)$.
 The intercepts of the red and pink trajectories with horizontal axis match well the initial position 
 of the basin boundaries, where the tracers split.
\label{fig: Zsqrflow}} 
 \end{figure}

As outlined above, numerical integration can be performed either directly in variables $\varphi,\theta$
or in WS variables $\zeta,\eta$ with additional transformation at each integration step from
the constants $\psi_j$ to the phases $\varphi_j$, to calculate the mean-field $Z$. Both methods match to a very good accuracy.  
Two examples of the time evolution shown in Fig. \ref{fig: Zsqrflow} for two 
random initial conditions ($N = 20$) illustrate this. 

In Fig. \ref{fig: Zsqrflow}, we highlight the trajectories of the two tracers that end up exactly at
asymptotic basin boundaries
$\tilde\theta_1 = \Phi(t=T_{\text{sync}})/2 + \pi/2$ and $\tilde\theta_2 = \Phi(t=T_{\text{sync}})/2 - \pi/2$ 
in $\varphi$-space (corresponding to pole of $\mathcal{M}_2$ in $\vartheta$-space 
at the final synchronous state), where $T_{\text{sync}}$ is the time at which some synchronization
threshold is reached during integration process. These trajectories are time-varying basin boundaries.  
This variation in time of basin boundaries is typically the case not just for higher-order coupling
like $\sim Z^2$, but also for the standard Kuramoto model.
These variations make it impossible 
to predict the initial locations of the basins, and therefore also unable to predict 
the numbers of oscillators in the final two clusters explicitly
from the initial condition alone. 
Because the basin boundaries being unstable trajectories in reverse time become attractive,
their positions at $t=0$ can be obtained by integrating back in time 
(under the correct mean-field time evolution
calculated forward in time), starting from any
point on the circle outside a small neighborhood 
from the two poles of $\mathcal{M}^{-1}_2$. The size
of the neighborhood $\epsilon \rightarrow 0$ under infinite forward integration time. Alternatively, 
we can simply map the singular constant $\tilde\psi$ via Eq.~\eqref{eq:bb} with $\zeta(0)$ and $\eta(0)$
as transformation parameters to obtain basin boundaries $\tilde\varphi$ at $t=0$. However, $\tilde\psi$
can only be known after integrating to full synchrony: $\tilde\psi=\Phi(\infty)-\eta(\infty)+\pi$.
Therefore, both methods of determining basin boundaries at $t = 0$ require
integration.  

\subsection{Comparison of the asymmetrical clustering under $Z^2$-mean-field model: prediction and numerics}
\begin{figure}
 \includegraphics[width=.45\textwidth]{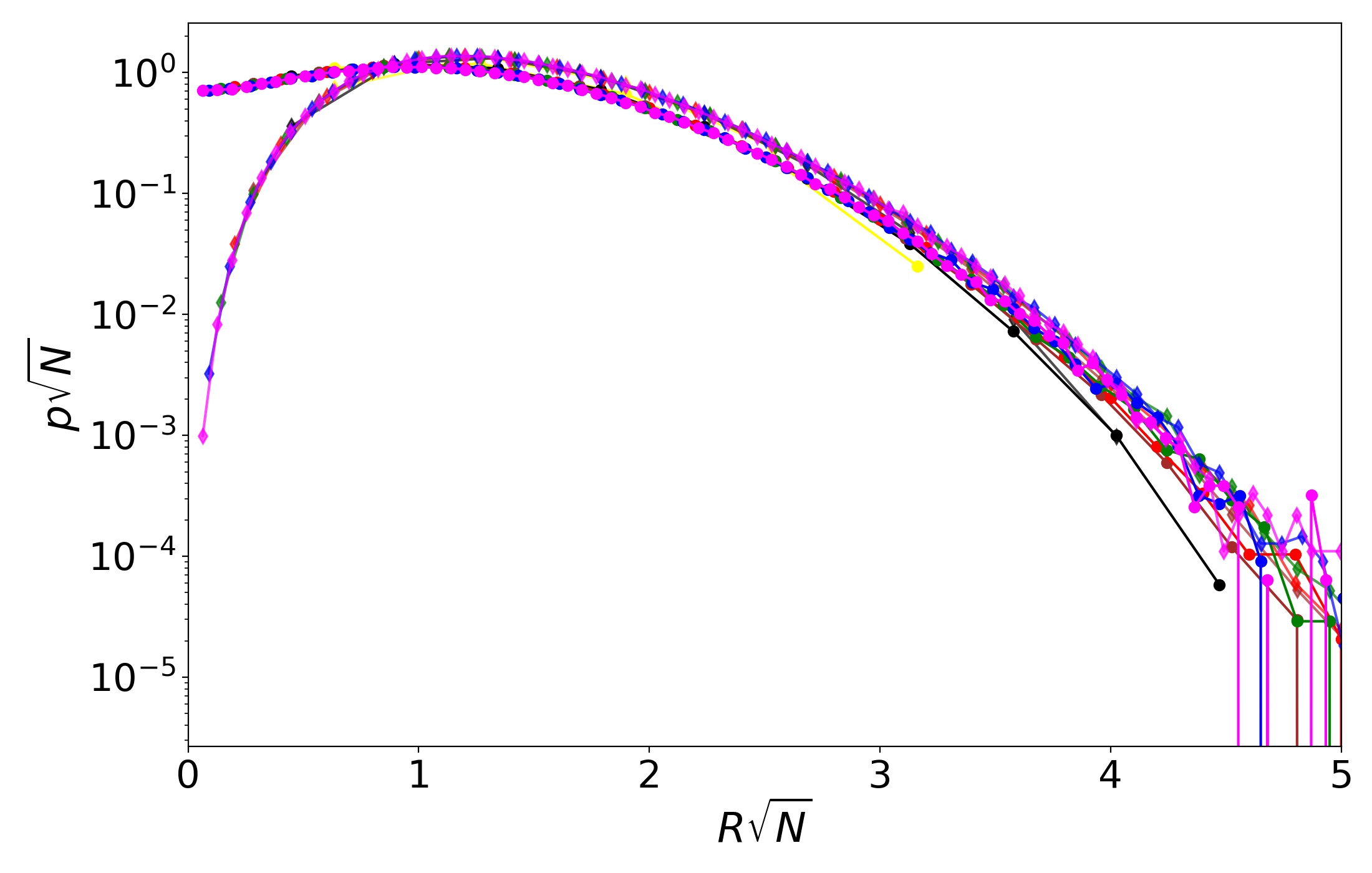}
 \caption{Comparison between prediction and simulation of the population sizes 
 of the two clusters, plotted as a histogram
 of $R$ values, based on random initial conditions (uniform distribution on a circle)
 for ensemble sizes $N = 10, 20, 50, 100, 200, 500, 1000$ (color-coded for $N$).
 Round markers: predictions of the cluster size based on 
 the basins estimated from initial states. Diamond markers:
 simulated results at steady state (data obtained from Ref.~\cite{maxim} with permission of the author).
 The distributions nearly coincide for $R\sqrt{N}\gtrsim 1$, but for small $R$ the estimated distribution
 does not reproduce the drop in the density near the totally absent symmetrical clustering state at $R=0$. 
\label{fig: comparison}} 
 \end{figure}

Here we discuss a way of approximating cluster distribution just from initial data. 
As discussed above, the basin boundaries rotate in the course of evolution.
 However, this rotation is 
usually small, which means we could estimate roughly the boundaries
using the initial value of $\zeta$ (according to our choice of initial condition). 
This method will therefore naturally involve an error corresponding to the degree of rotations. 
Using the same expression as the final singular points $\tilde\varphi: \Phi(\infty)/2 + \pi/2$ and 
$\Phi(\infty)/2 + 3\pi/2$, we 
approximate basin boundaries at initial time as
 $\Phi(0)/2 + \pi/2$ and $\Phi(0)/2 + 3\pi/2$.
 Since the initial condition is
$\zeta(0) = H(0) = Z^2(0)$, this implies $\arg(\zeta(0))=\Phi(0) = 2\arg[Z(0)]$.
The number of oscillators falling into each basin (marked by $\arg(Z) + \pi/2$
and $\arg(Z) + 3\pi/2$ at $t=0$)
yields therefore an estimate for populations of the final clusters. 
In Fig. \ref{fig: comparison}, this estimation in the form of probability distribution 
is compared with the correct final asymmetrical clustering distribution, 
as a function of the metric $R = |Z|$, both axes scaled by $\sqrt{N}$. 

This estimate is naturally not accurate because of the rotation of the basin boundaries, 
however, it is able to explain several features
 of the distribution. First, the asymmetry of the distribution, i.e., 
 the maximum of the distribution is not at $R=0$ (the symmetric clustering state), is present
 for both the prediction and simulation. However, the location of the maximum is underpredicted
 by theory. Second, the $\sqrt{N}$ scaling law with respect to the ensemble size applies to both. 
 In fact, the successful scaling of the prediction based only on initial conditions
 implies that the source of the steady state scaling law lies in 
 the initial condition and their finite sampling, not in the dynamics. 
 
 According to the first observation, the source of this ``symmetry breaking'' in terms of particle distribution
 should in part be related to the geometrical fact that the angle of the particle 
 mean-field $\arg[Z(0)]$ is not isotropic on 
 the circle, even if the underlying particle distribution is isotropic on average. 
 Because if $\arg[Z(0)]$ is isotropic, then we should see a distribution more akin to the binomial distribution.
 This is intuitive when one reflects on the meaning of $\arg[Z(0)]$ as the direction of 
 the average over-density of the initial phase distribution. There will be by definition more
 phases on the side where $\arg[Z(0)]$ is pointing towards, and fewer on the side
 opposite to it. Naturally, the half circle spanning these two sides marked by $\arg[Z(0)]+\pi/2$
 and $\arg[Z(0)]+3\pi/2$ will have unequal number of phases in these ``approximate basins''. 
 However, our estimate is completely reliant on our choice of initial conditions for the 
 WS variables. The choice is arbitrary due to over-determinedness of the transformed equations. 
 Therefore, another choice of initial conditions will give another completely different estimate.
 The fact that our estimate seems to explain some features of the final distribution speaks only
 for the ``naturalness'' of our choice of initial conditions, justified by the WS equation \eqref{eq:wsii_realimag}. 
 Conversely, another cleverer choice might be able to exactly predict 
 the final distribution, if such a choice exists for all initial conditions. 
 
 Despite the partial explanation for the final asymmetrical clustering, 
 the estimate fails to predict the lack of states near $R=0$, as well as 
 the complete absence of the symmetrical state (two clusters being equally sized). 
 This failure can only be due to the dynamics of the system, which is not inferrable
 directly from the initial conditions, even though the system is fully deterministic. 
 Specifically, in simulations, the $R=0$ final state is completely absent, which is
 in fact due to the weak instability at the symmetry state. An elementary linear stability analysis 
 of the symmetry states with $N=2$ or $N=4$
shows that the states (two clusters with sizes 1-1 or 2-2) are weakly unstable, 
thus giving evidence of the weak instability at the symmetrical state, justifying
their absence from the distribution. 

\subsection{Possibility for decreasing mean-field in the $Z^2$-mean-field model under positive coupling}
\label{sec:dmf}
\begin{figure}[h]
\centering 
 \includegraphics[width=0.4\textwidth]{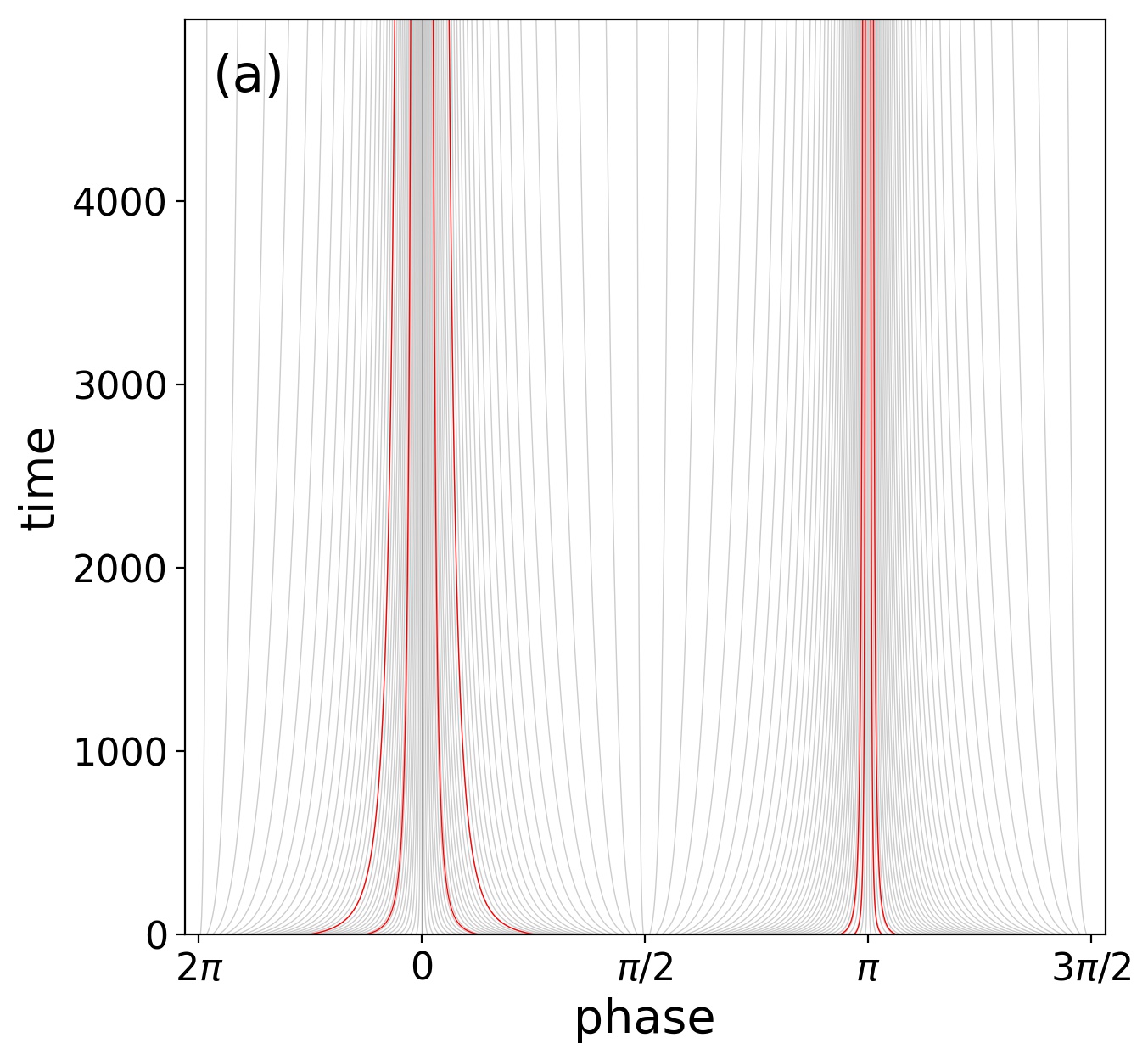} 

 \includegraphics[width=0.4\textwidth]{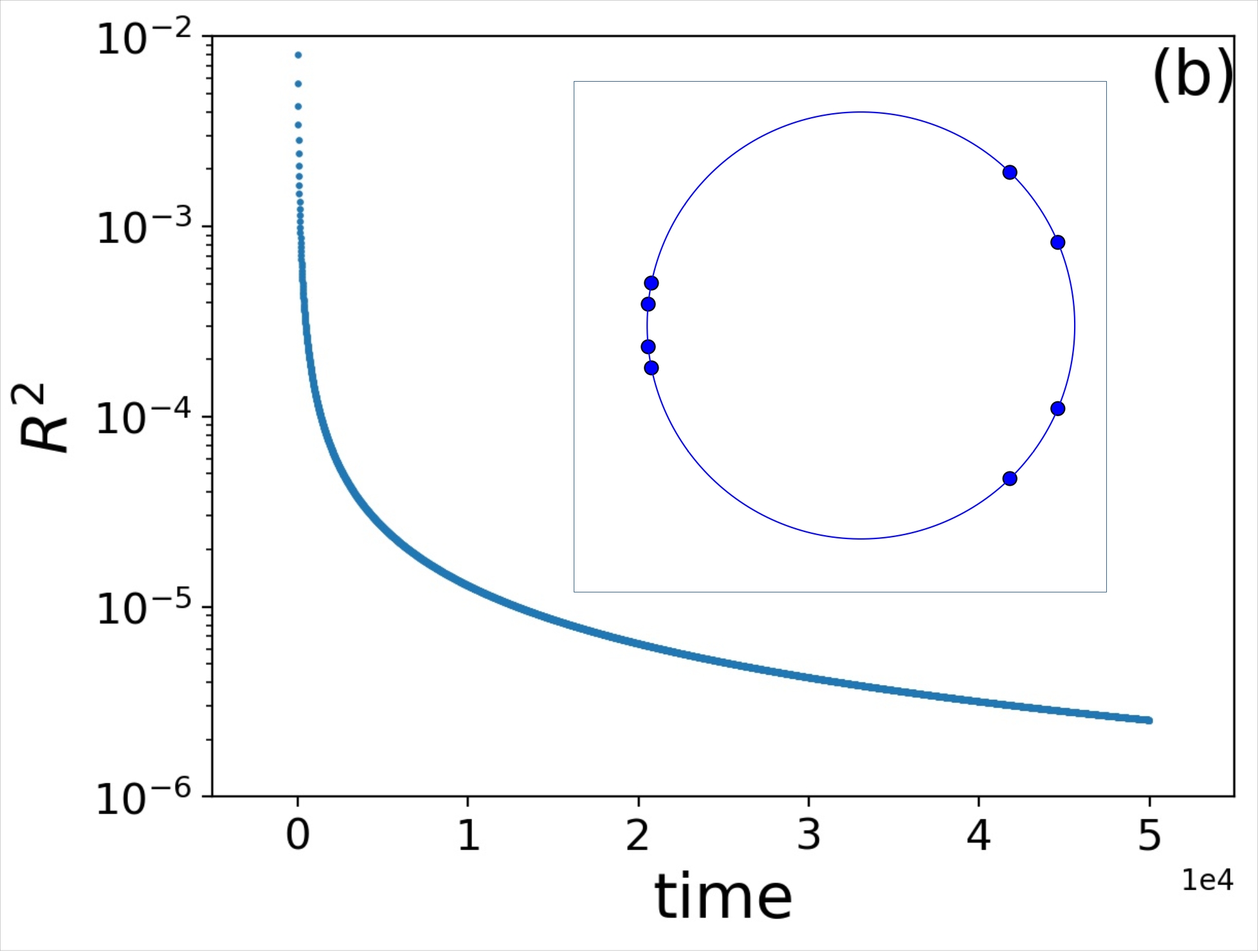}
 (a) \hspace{3.5cm} (b) 
 \caption{(a): Flow of passive (gray) and active oscillator phases (red). (b): Evolution of the mean-field amplitude $R^{2}$.
  Inset in panel (b): Special symmetric initial conditions.}
 \label{fig:tricking}
\end{figure}

The Kuramoto model with first-order mean-field coupling is known
to possess a Lyapunov function~\cite{ws94}. This means that generic
initial conditions (i.e., with an initially non-zero order parameter)
monotonously evolve towards a synchronous clustered state under attractive coupling
(only initial states with vanishing mean-field do not evolve). This property
is not shared by the $Z^2$ second-order coupling model we consider here. 
It is possible, using symmetry, to construct special initial conditions
which lead to a monotonic decrease of the order parameter. 
For example, we consider special symmetric eight initial phase values as shown 
in Fig. \ref{fig:tricking}(b) inset.
The initial value of the Kuramoto 
order parameter is nonzero, $R>0$, i.e., the system will evolve under $Z^{2}$.
However, the evolution preserves the symmetry, so a formation of
asymmetric clusters is not possible.
Numerical integration shown in Figure \ref{fig:tricking}(a) demonstrates convergence
towards an unstable configuration with $R=0$. 
One should note, that numerical errors could eventually destabilize this symmetric 
state due to symmetry breaking, with a formation of two clusters with 
sizes 5 and 3 each, which should eventually be observed on a long timescale.

\section{Discussion and Conclusion} \label{sec:summary}
Our study provides an analytical extension of the dimension-reducing formulation 
of globally coupled identical phase oscillators under pure higher-order harmonic coupling,
and carries the analytical tradition of Watanabe-Strogatz theory further, the same way 
Ref. \cite{Skardal} did in terms of the OA theory for the Kuramoto model. 
Similar to the WS formulation for a first harmonic
coupling, we apply an analogous type of M{\"o}bius transformation from the space of the original phases 
into the space of the transformed phases (constants of motion) to obtain the three-dimensional 
WS equations. We devise an argument to solve the apparent non-unique transformation
from the constants back to the original phases.
Numerical integration shows that the simulation based on reduced WS equations matches 
the simulation based on the phase equations. 

As an example, the WS formulation of the $Z^2$-mean-field model, which exhibits
asymmetrical clustering, is tested with good numerical agreement to the phase model.
The  boundaries of the basins of attraction under such a model match the pole in the M{\"o}bius map at 
the final steady state. The asymmetric clustering can be explained, albeit partially via the
theory, explicitly from the initial distribution of phases. The main obstacle is the fact 
that the pole only appear in the M{\"o}bius map at the final synchronous steady state,
and not at intermediate and initial states. This makes it impossible to find the 
initial basin boundaries without following the dynamics to the final state.

We also report on a possibility for (unstable) desynchronization to happen in the attractively coupled
$Z^2$-mean-field model, a situation not observed in the classic Kuramoto setup.
This is an indication for complex non-monotonous transient behaviours in identical
ensembles with higher-order coupling.  

Currently, both WS and OA formulations are limited to pure $l$-harmonic coupling, and are
not applicable to mixed harmonics coupling.
 Besides this constraint on the form of the coupling, 
these approaches are restricted also by the connection topology (global coupling,
or its modifications like star graph~\cite{stargraph}, is usually required),
and by the natural frequency distribution of the oscillators (identical in the case
of WS, Cauchy in the case of OA). It appears promising to extend the WS and OA
theories via perturbation analysis, first attempts in this direction have been 
reported recently~\cite{Vlasov-Rosenblum-Pikovsky-16,Tyulkina_etal-18}.

\begin{acknowledgments}
This paper is developed within the scope of the IRTG 1740/TRP 2015/50122-0, funded 
by the DFG/ FAPESP. Work of A.P. is supported by Russian Science Foundation (study of a hypernetwork: Grant Nr.
17-12-01534, development of numerical approach: Grant Nr. 19-12-00367).
\end{acknowledgments}

\section{Appendix} \label{Appendix}
\begin{figure}[h!]
\includegraphics[width=0.9\columnwidth]{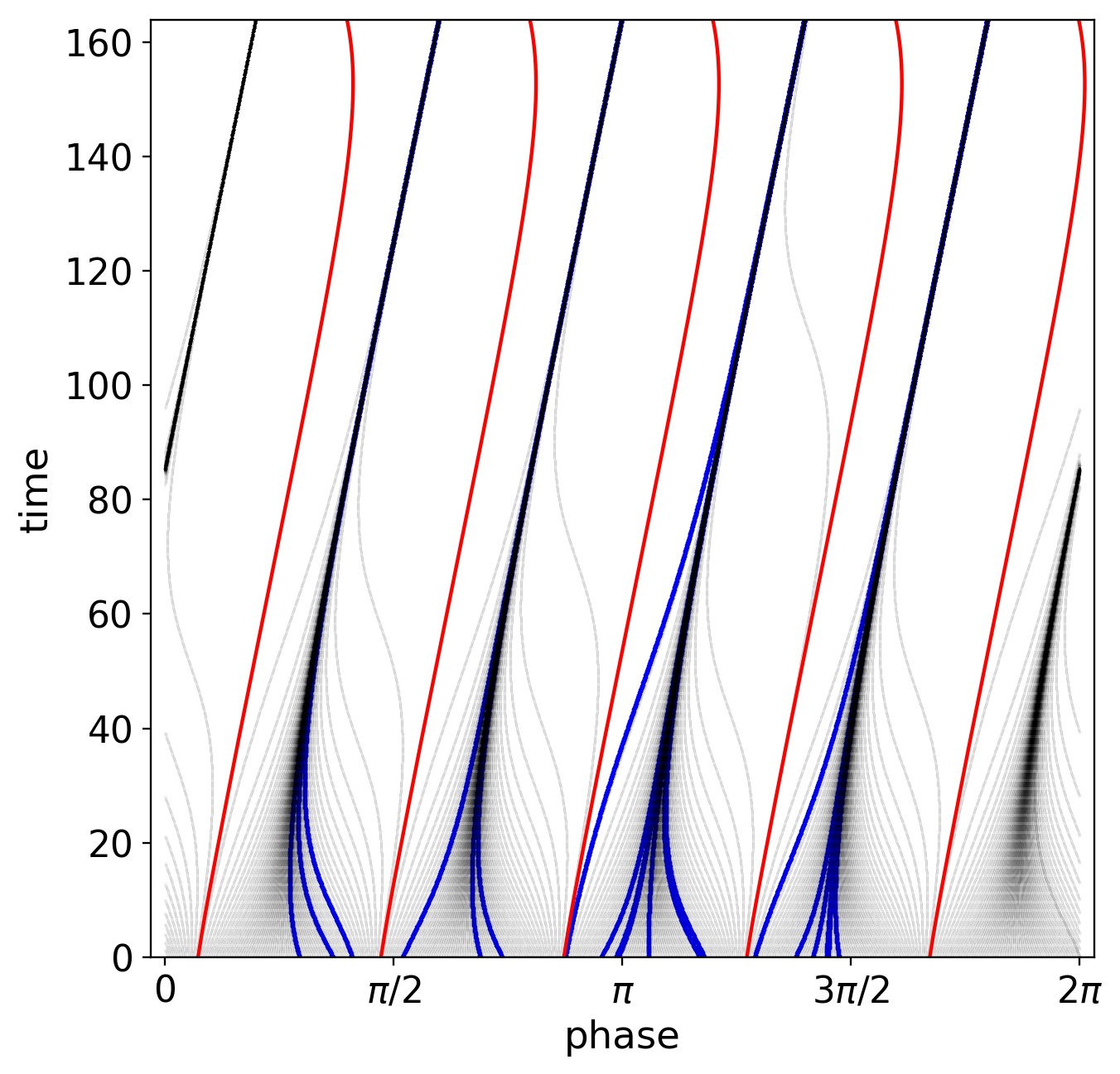}
	\caption{Analogous to the $Z^2$ case in Fig. \ref{fig: Zsqrflow}, 
	here a flow plot for model $\dot \varphi_{j} =  \operatorname{Im}[ Z^5 e^{-i5\varphi_{j}}]$, 
	for $N = 20$. Red curves are the basin boundaries $\tilde\varphi(t)$.
	}
	\label{fig:5th}
\end{figure}

Since the theory proposed in this paper --- that phase variables with identical frequency globally coupled
via pure higher-order coupling is partially integrable --- is in fact valid at any order $l$, we
provide in Fig.~\ref{fig:5th} an example where $l=5$ for 20 oscillators with random uniform initial conditions. 

Analogous to Fig. \ref{fig: Zsqrflow}, where a second-order example is provided, for $l = 5$ and a forcing term  
$H = Z^5$ where $Z$ is the Kuramoto mean-field, Fig. \ref{fig:5th} shows the phase flow plot (along with passive tracers). 
As with the $H= Z^2$ model mentioned in the main text, we can find the basin boundaries of cluster formation numerically. 
Also note that it is generally possible to arrive at a number of clusters smaller than 
the order of coupling which gives the maximal number of clusters; in this case four clusters under fifth-order coupling.



\bibliographystyle{unsrt}
\bibliography{biblio}

\end{document}